\documentclass[pra,showpacs,preprint,superscriptaddress]{revtex4}
\usepackage{amsmath}
\usepackage{amsfonts}
\usepackage{graphicx}
\usepackage{longtable}
\newcommand{\be}{\begin{equation}}
\newcommand{\ee}{\end{equation}}

\newcommand{\ba}[1]{\left(\begin{array}{#1}}
    \newcommand{\ea}{\end{array}\right)}

\begin{document}
        \title{Spin squeezing in Dicke-class of states with non-orthogonal spinors}
				
				\author{K. S. Akhilesh}
\affiliation{Department
            of Studies in Physics, University of Mysore, 
            Manasagangotri, Mysuru-570006, Karnataka, India }
\author{K. S. Mallesh}
\affiliation{Department 
            of Studies in Physics, University of Mysore, 
            Manasagangotri, Mysuru-570006, Karnataka, India }
\author{Sudha}
\email{arss@rediffmail.com}
\affiliation{Department of Physics, Kuvempu University, Shankaraghatta, Shimoga-577 451, India.}
\affiliation{Inspire Institute Inc., Alexandria, Virginia, 22303, USA.}
\author{Praveen G. Hegde} 
\affiliation{Department 
            of Studies in Physics, University of Mysore, 
            Manasagangotri, Mysuru-570006, Karnataka, India }
 \date{\today}

        \begin{abstract}
        The celebrated Majorana representation is exploited to 
        investigate spin squeezing in different classes of pure symmetric 
        states of $N$ qubits with two distinct spinors, namely the Dicke-class of states. On obtaining a general 
        expression for spin squeezing parameter, the variation of squeezing for different configurations is studied in detail. It is 
				shown that the states in the Dicke-class, characterized by two-distinct non-orthogonal spinors, exhibit squeezing.   
        \end{abstract}
				\pacs{03.67.Mn}
	\maketitle				
        \section{Introduction}
        
        Spin, an intrinsic degree of freedom of the physical systems 
        is a fascinating area of study as it is entirely of quantum origin not having a classical analogue. 
				The components $\hat{S}_x$, $\hat{S}_y$, $\hat{S}_z$ of the spin operator 
				$\hat{S}$ are non-commuting and hence obey the uncertainty relation 
        \begin{equation} 
        (\Delta {\hat{S}}_{x})^{2} (\Delta {\hat{S}}_{y})^{2} \geq 
				\frac{ 
            \vert \langle {\hat{S}}_{z}\rangle\vert^{2}}{4}.
        \end{equation} 
	Based on the uncertainty relation satisfied by position, momentum operators, the concept of squeezing was first introduced for the states 
        of harmonic oscillator and it was later extended to the radiation 
        fields~\cite{kim1,kim2,kim3}. A 
        comparison of the uncertainty relation satisfied by the components of spin operator 
				$\hat{S}=(\hat{S}_x,\,\hat{S}_y,\,\hat{S}_z )$  with that of the  
        non-commuting operators of bosonic systems led to the 
        extension of the concept of squeezing to spin systems 
        ~\cite{ku}. In the case of spin systems the 
        lower bound which is essential for the study of squeezing 
        needs a careful consideration because of its coordinate 
        dependence. This led to the conclusion that the 
        behaviour of squeezing is dependent on the choice of the coordinate frame. 
        Kitegawa and Ueda~\cite{ku} showed that the occurrence of 
        squeezing in spin $s$ state is due to the existence of 
        quantum correlations among the constituent $2s$ spin $1/2$ systems 
        and 
        suggested a 
        coordinate 
        independent definition of spin squeezing. According to Kitegawa-Ueda~\cite{ku}, a $N$-qubit state is KU spin-squeezed if 
the minimum variance of a spin component normal to the {\emph{mean spin direction}} is smaller than the 
standard quantum limit $\frac{\sqrt{N}}{2}$ of the spin coherent 
state. Thus, for a spin squeezed state, 
\[
(\Delta {\hat{S}}_\bot)_{\mbox{min}} \leq \frac{\sqrt N}{2} \Longrightarrow \frac{2(\Delta {\hat{S}}_\bot)_{\mbox{min}}}{\sqrt{N}}\leq 1, 
\]
and hence the Kitegawa-Ueda spin squeezing parameter  is defined as, 
\[
\xi=\frac{2(\Delta {\hat {S}}_\bot)_{\mbox{min}}}{\sqrt{N}}.
\]
$\xi$ is a quantitative measure of spin squeezing and for a spin squeezed state  $0\leq \xi<1$. 

Spin squeezing has been an intense area of study~\cite{ku,wn2,gsapuri,puri,msd,sor,ulamku,um,usha1,usha2,s1,s2,us1,us2,dyan,akru,gtoth,Toth2011,ReviewNori,song,divsu} due to its theoretical importance as well as the applicability of spin squeezed states in the fields of low noise, high precision spectroscopy~\cite{caves,yurke,kunew2} and quantum information science~\cite{sor,ulamku,usha1,usha2,s1,s2}. The relative ease with which spin squeezed states can be produced~\cite{sor}, has made them more accessible and hence useful in the concerned fields.

In 1932, Ettore Majorana ~\cite{maj} had shown that a pure state with spin $s=N/2$ has one to one correspondence with the 
symmetrized combination of pure states of $N$-spin $1/2$ systems ($N$-spinors) and provided an elegant geometrical representation of a symmetric multiqubit state as $N$ points on the Bloch sphere. Based on the number of distinct spinors (diversity degree) and the frequency of their occurrence (degeneracy configuration), a useful classification of pure symmetric multiqubit states into SLOCC
(stochastic local operations and classical communication) inequivalent families has been achieved~\cite{bas,usha4}. Recently, permutation symmetric states have received a revived attention in diverse branches of physics due to their geometrical intuition and experimental importance~\cite{sac,usha1,usha2,usha3}.

Dicke states $\vert s,\,m\rangle$, the common eigenstates of $\hat{S}^2$, $\hat{S}_z$ belonging respectively to eigenvalues $s(s+1)\hbar^2$,  
$m\hbar$, $m=-s,\,-s+1,\,\ldots,s-1,\, s$, are pure symmetric states consisting of two distinct {\emph {orthogonal}} spinors $\vert 0 \rangle$, $\vert 1 \rangle$. The different degeneracy configurations exhaust all Dicke states and each degeneracy configuration corresponds to an SLOCC inequivalent class~\cite{bas}. It is well known that Dicke states exhibit no spin-squeezing inspite of them being entangled states{\footnote{In fact, spin-squeezing implies pairwise entanglement in multiqubit symmetric states but the converse is not always true.}}.   It would therefore be of interest to examine the spin squeezing nature of $N$ qubit symmetric states consisting of all possible permutations of two distinct non-orthogonal spinors, the natural extension of Dicke states. The class of $N$-qubit pure symmetric states with two distinct spinors (diversity degree $2$) are referred to as {\emph {Dicke-class of states}, with the Dicke states corresponding to symmetrized combination of two orthogonal spinors 
$\vert 0 \rangle$, $\vert 1\rangle$. An analysis of 
spin squeezing in the Dicke-class of states with two distinct non-orthogonal spinors, taking into account all possible degeneracy configurations among them, is the aim of this work. In view of the connection between spin squeezing and pairwise entanglement in multiqubit states~\cite{ulamku,usha1,usha2,s1,s2}, this study will be quite useful in quantum information science.  

This paper is organized as follows: The first section gives an introduction to the concept of spin squeezing and provides a brief overview of Majorana representation of pure symmetric multiqubit states. On defining the Dicke-class of states in Section I, the structure of these states in their canonical form is given in Section~II. The mean spin vector of each SLOCC inequivalent family of Dicke-class of states is identified in Section~II. Among the spin components perpendicular to the mean spin vector, the one which gives minimum variance is identified in Section~III for all degeneracy configurations of the Dicke-class of states. Using the results of Section II and Section III, the spin squeezing parameter is explicitly given in Section~IV. Illustrative graphs of spin squeezing parameter for several different degeneracy configurations (corresponding to different SLOCC classes) are given in Section~IV, each of them indicating spin squeezing in Dicke-class of states with two distinct non-orthogonal spinors. The amount of spin squeezing in each SLOCC class, as a function of number of qubits is discussed in detail 
(Section~IV). The concluding section (Section~V) summarizes the results of the paper and highlights their significance.

\section{Canonical form of Dicke-class of states and identification of their mean spin vector}
        
A pure symmetric state of $N$ qubits consisting  of two spinors, spinor $\vert u_1\rangle$ appearing $k$ times and  spinor 
        $\vert u_2\rangle$ appearing $N-k$ times, is given in the Majorana representation~\cite{maj} by
\begin{equation}
\label{maj1}
        \vert{\Psi_{k,\,N-k}}\rangle =\frac{1}{\mathcal{N}}\sum \limits_{P} [ 
        \underbrace{\vert{u_{1}}\langle \otimes \vert{u_{1}}\rangle \otimes  \ldots 
            \otimes  \vert{u_{1}}\rangle}_{k} \otimes \underbrace{\vert{u_{2}}\rangle 
            \otimes \ldots  \vert{u_{2}}\rangle}_{N-k}]. 
        \end{equation}
				Here $P$ denotes all possible permutations of the two spinors and 
        $\cal{N}$ 
        is the normalization factor.  Without any 
        loss of 
        generality, one can employ a coordinate system in 
        which z-axis is directed along one of the two qubits and x-z plane contains the other. 
				On choosing $\vert{u_{1}}\rangle=\ba{c} 1 \\ 0 \ea$, it readily follows that 
        $\vert{u_{2}}\rangle=\ba{c} a \\ \sqrt{1-a^{2}} \ea$, $0\leq a < 1$ being real. The canonical form of Dicke-class of states is thus equivalent to the state $\vert{\Psi_{k,\,N-k}}\rangle$ and we have  
				\begin{equation}
        \vert{\Psi_{k,\,N-k}}\rangle\equiv \frac{1}{\mathcal{N}}\sum \limits_{P} [ 
        \underbrace{\vert{0}\rangle \otimes \vert{0}\rangle \otimes  \ldots 
            \otimes  \vert{0}\rangle}_{k} \otimes \underbrace{\vert{u_{2}}\rangle 
            \otimes \ldots  \vert{u_{2}}\rangle}_{N-k}], \ \ \  \vert{u_{2}}\rangle=\ba{c} a \\ \sqrt{1-a^{2}} \ea, \ \ 0\leq a < 1, 
        \end{equation}
				as the canonical form of Dicke-class of states. One can see that $a=0$ leads 
				to the spinor $\vert u_2\rangle=\vert 1 \rangle$, which is orthogonal to $\vert u_1\rangle=\vert 0\rangle$. The Dicke states are thus given in Majorana representation by 
				\begin{equation}
				\label{majcan}
        \vert{\Psi_{k,\,N-k}}\rangle\equiv \frac{1}{\mathcal{N}}\sum \limits_{P} [ 
        \underbrace{\vert{0}\rangle \otimes \vert{0}\rangle \otimes  \ldots 
            \otimes  \vert{0}\rangle}_{k} \otimes \underbrace{\vert{1}\rangle 
            \otimes \vert{1}\rangle \otimes \ldots  \vert{1}\rangle}_{N-k}] ,  
        \end{equation}
	The non-zero values of $a$ in $\vert{u_{2}}\rangle=\ba{c} a \\ \sqrt{1-a^{2}} \ea$, i.e., $0<a<1$ lead to the Dicke-class of states 
	$\vert{\Psi_{k,\,N-k}}\rangle$ with non-orthogonal spinors. When $a=1$, we have $\vert u_2\rangle=\vert 0 \rangle$ which leads to the separable state (corresponding to $k=N$) 
	$\vert{\Psi_{N,\,0}}\rangle=\vert{0}\rangle \otimes \vert{0}\rangle \otimes \ldots \otimes \vert{0}\rangle \otimes \vert{0}\rangle$	and is not of interest as far as its quantum correlation features are considered. 
				
In order to evaluate the spin squeezing parameter for the states $\vert{\Psi_{k,\,N-k}}\rangle$, one needs to identify the mean spin direction $\vert \hat{n}_0\rangle$ and a direction $\vert \hat{n}_\bot\rangle$ perpendicular to $\vert\hat{n}_0\rangle$ which gives minimum variance $\Delta {\hat{S}}_\bot$ of the spin operator $\hat{S}$,  for all values of $N$ and $k<N$. It is worth recalling here that a unit vector along the mean spin direction of any multiqubit symmetric state is given by 
\begin{equation}
{\hat n}_0=\frac{\left(\langle\hat{S}_x\rangle,\,\langle\hat{S}_y\rangle,\,\langle\hat{S}_z\rangle\right)}{\sqrt{\langle\hat{S}_x\rangle^2+\langle\hat{S}_y\rangle^2+\langle\hat{S}_z\rangle^2}}, \ \ {\hat{S}_i}=\frac{1}{2} \sum_{k=1}^N \, \hat{\sigma}_i^{(k)}, \ \ i=x,\,y,\,z
\end{equation} 
where $\hat{\sigma}_i^{(k)}$ are the Pauli spin operators corresponding to $k$th qubit. The expectation values of the spin operators are evaluated with respect to the state $\vert{\Psi_{k,\,N-k}}\rangle$ in Eq. (\ref{majcan}) i.e., $\langle\hat{S}_i\rangle=
\langle \Psi_{k,\,N-k}\vert {S}_i\vert \Psi_{k,\,N-k}\ \rangle$, $i=x,\,y,\,z$. 

In order to find out the mean spin direction ${\hat n}_0$, we need to evaluate the expectation values $\langle\hat{S}_x\rangle$, $\langle\hat{S}_y\rangle$ and $\langle\hat{S}_z\rangle$. As the spinor $\vert u_1\rangle=\vert 0\rangle$ is associated with the unit vector $(0,\,0,\,1)$ along z-axis and $\vert u_2\rangle=\ba{c} a \\ \sqrt{1-a^2} \ea$ corresponds to unit vector $\left(2a\sqrt{1-a^2},\,0,\,2a^2-1\right)$ in the x-z plane, it is not difficult to see that the expectation value $\langle\hat{S}_y\rangle$ is zero. On explicit evaluation also, we have verified that 
$\langle\hat{S}_y\rangle=0$ for all degeneracy configurations. The expectation values $\langle\hat{S}_x\rangle$, 
$\langle\hat{S}_z\rangle$ are explicitly evaluated for all possible values of $k$ for each $N$ from $N=2$ to $N=5$ and are listed in Tables \ref{table01}, 
\ref{table02} respectively.
 
\begin{center} 			
\begin{table}[ht] 
\caption{$\langle \hat{S}_{x}\rangle$ for different degeneracy configurations} 
\label{table01} 
\begin{tabular}{|c|l|} 
\hline 
$N=2,\,k=1$ & $ \frac{1}{(1+a^{2})} \left[ 2a\sqrt{1-a^{2}}\right] $ \\ 
            \hline
            $N=3,\,k=2$  & $\frac{1}{(1+2a^{2})} \left[ 
            3a\sqrt{1-a^{2}} \right]$ 
            \\ \hline
            $N=3,\,k=1$  & $\frac{1}{(1+2a^{2})} \left[ 
            2a\sqrt{1-a^{2}}(2+a^{2}) \right]$ 
            \\ \hline
            $N=4,\,k=3$ & $\frac{1}{(1+3a^{2})} \left[ 
            4a\sqrt{1-a^{2}} \right]$ \\ \hline
            $N=4,\,k=2$ & $\frac{1}{(1+4a^{2}+a^{4})} \left[ 
             6a\sqrt{1-a^{2}}(1+a^{2})
            \right]$ \\ \hline
            $N=4,\,k=1$ & $\frac{1}{(1+3a^{2})} 
            \left[ 
            6a\sqrt{1-a^{2}}(1+a^{2})\right] $ \\  \hline
             $N=5,\,k=4$ & $\frac{1}{(1+4a^{2})} 
            \left[ 
            5a\sqrt{1-a^{2}}\right] $ \\ 
            \hline
            $N=5,\,k=3$ & $\frac{1}{(1+6a^{2}+3a^{4})} 
           \left[ 
           4a\sqrt{1-a^{2}}(2+3a^{2})\right] $ \\ 
            \hline             
            $N=5,\,k=2$ & $\frac{1}{(1+6a^{2}+3a^{4})} 
            \left[ 
            3a\sqrt{1-a^{2}}(3+6a^{2}+a^{4})\right] $ \\  \hline
             $N=5,\,k=1$ & $\frac{1}{(1+4a^{2})} 
            \left[ 
            4a\sqrt{1-a^{2}}(2+3a^{2})\right] $ \\  \hline
           \end{tabular}
\end{table} 
\end{center}
\begin{table}[ht]
\caption{$\langle{ 
            \hat{S}_{z}}\rangle$ for different degeneracy configurations}
\label{table02}						
\begin{tabular}{|c|l|}
\hline
            $N=2,\,k=1$ & $ 
            \frac{`1}{(1+a^{2})}[2a^{2}]  $ \\ 
            \hline
            $N=3,\,k=2$  & $\frac{1}{2(1+2a^{2})}[1+8a^{2}] $ 
            \\ \hline
            $N=3,\,k=1$  & $\frac{1}{2(1+2a^{2})}[4a^{4}+6a^{2}-1] $ 
            \\ \hline
            $N=4,\,k=3$ & $\frac{1}{(1+3a^{2})} [1+7a^{2}]$ \\ 
            \hline
            $N=4,\,k=2$ & $\frac{1}{(1+4a^{2}+a^{4})} [6a^{4}+6a^{2}] $ \\ 
            \hline
            $N=4,\,k=1$ & $\frac{1}{(1+3a^{2})} [6a^{4}+3a^{2}-1] $ \\ 
            \hline
            $N=5,\,k=4$ & $\frac{1}{2(1+4a^{2})}[3+22a^{2}] 
            $ \\ 
            \hline
            $N=5,\,k=3$ & $\frac{1}{2(1+6a^{2}+3a^{4})}[1+22a^{2}+27a^{4}] 
            $ \\ \hline
            $N=5,\,k=2$ & 
            $\frac{1}{2(1+6a^{2}+3a^{4})}[6a^{6}+30a^{2}+45a^{4}+1] 
            $ \\ \hline
            $N=5,\,k=1$ & 
            $\frac{1}{2(1+4a^{2})}[4a^{2}+24a^{4}-3] 
            $ \\ \hline
    \end{tabular}
\end{table} 

On a careful observation of the expectation values listed in tables \ref{table01}, \ref{table02}, a general expression for  
$\langle \hat{S}_x\rangle$, $\langle \hat{S}_y\rangle$ applicable for all values of $N$ and $k<N$ can be obtained, and they are given by 
 
\begin{eqnarray}\nonumber
	\langle \hat{S}_{x}\rangle & = & \frac{N 
	a\sqrt{1-a^{2}}}{{\cal{N}}^{2}}\left\{ 
	\frac{1}{2}\ba{c} {N-1} \\ 
	{N-k} \ea \sum_{r=0}^{N-k} \ba{c}{k-1}\\ {r}\ea 
	\ba{c}{N-k}\\ {r+1}\ea a^{2r} \right.\nonumber \\ 
	& & \left. + \ba{c}{N-1}\\ {N-k-1}\ea \sum_{r=0}^{N-k-1}\,a^{2r}
\ba{c}{N-k-1}\\{r}\ea \left[ \frac{1}{2} \ba{c}{k}\\ {r+1}\ea  + \ba{c}{k}\\ {r}\ea \right] \right\} \\ \nonumber
& & \\ \nonumber 
\langle \hat{S}_{y}\rangle & = & 0 \nonumber \\ 
& &  \nonumber \\
\langle \hat{S}_{z}\rangle & = & \frac{N}{2{\cal{N}}^{2}}\left\{ \ba{c}{N-1}\\ {N-k} \ea
\sum_{r=0}^{N-k}\, a^{2r}\,\ba{c}{k-1}\\ {r}\ea \left[ 
 \ba{c} {N-k} \\ {r} \ea   + a^{2}  \ba{c}{N-k}\\ {r+1}\ea \right] \right. \\  
& & \left. +\ba{c}{N-1}\\ {N-k-1}\ea  \sum_{ 
r=0}^{N-k-1} a^{2r} \ba{c} {N-k-1}\\ {r} \ea \left[a^{2}
\ba{c}{k}\\ {r+1}\ea   +(2a^{2}-1) \ba{c}{k}\\{r}\ea\right]  
		\right\} \nonumber
\end{eqnarray}
where  $N$ is the number of 
qubits and ${\cal{N}}$ is the normalization constant given by
\[ {\cal{N}}^{2}= \ba{c}{N}\\ {k} \ea \sum_{r=0}^{N-k} \ba{c}{k}\\ {r}\ea 
\ba{c}{N-k}\\ {r} \ea a^{2r}.\]

Since $\langle \hat{S}_{y} \rangle=0$, the mean spin direction becomes,
\be
\label{n0}
\hat{n}_{0}=\frac{1}{\sqrt{\langle \hat{S}_x \rangle^2+\langle \hat{S}_z \rangle^2}}\left(\langle \hat{S}_x \rangle,\, 0,\, \langle \hat{S}_z \rangle\right) 
\ee
A plane perpendicular to $\hat{n}_{0}$ is defined by the two mutually orthogonal unit vectors  
given by 
\be 
\label{n1n2}
\hat{n}_1=(0,\,1,\,0) \ \  \text{and}\ \ \ 
\hat{n}_2=\frac{1}{\sqrt{\langle \hat{S}_x \rangle^2+\langle \hat{S}_z \rangle^2}}\left(-\langle \hat{S}_z \rangle,\, 0,\, \langle \hat{S}_x \rangle\right).
\ee
Any vector in the plane defined by $\hat{n}_{1}$, $\hat{n}_{2}$, are perpendicular to $\hat{n}_{0}$ and the task is to identify a unit vector 
say $\hat{n}_\bot$ such that the spin component $\hat{S}.\hat{n}_\bot$ has minimum variance.  

\section{Identification of the minimum variance $(\Delta \hat{S}_\bot)_{\text{min}}$}

We recall here the definition of spin squeezing adopted in~\cite{ku}. 
A pure symmetric state of $N$ qubits is said to be squeezed in a direction perpendicular to the mean spin direction, iff 
the spin squeezing parameter $\xi$  given by
\begin{equation}
\label{xi}
\xi=\frac{2 (\Delta \hat{S}_\bot)_{\text{min}} 
}{\sqrt{N}}.
\end{equation} 
lies between $0$ and $1$ ($0\leq \xi<1$).

Here $(\Delta {\hat{S}}_\bot)_{\mbox{min}}=\left(\Delta(\hat{S}\cdot\hat{n}_{\perp})\right)_{\text{min}}$ 
is the minimum uncertainty (variance) of spin angular momentum component in the 
plane perpendicular to the mean spin direction $\hat{n}_{0}$. As the vector $\hat{n}_\perp$ lies in the plane defined by 
vectors $\hat{n}_1$, $\hat{n}_2$ both of which are perpendicular to the mean spin direction $\hat{n}_0$, we denote 
$\hat{n}_\perp=\hat{n}_1 \cos \Phi +\hat{n}_2 \sin \Phi$. In view of the fact that $\langle \hat{S}\cdot \hat{n}_\perp \rangle=0$,
\[
(\Delta {\hat{S}}_\bot)^2= 
\left\langle \left(\hat{S}\cdot \hat{n}_\perp\right)^2 \right\rangle -\left\langle \hat{S}\cdot \hat{n}_\perp \right\rangle^2=\left\langle \left(\hat{S}\cdot \hat{n}_\perp\right)^2 \right\rangle.
\]
and minimizing $\left\langle \left(\hat{S}\cdot \hat{n}_\perp\right)^2 \right\rangle$ is equivalent to minimizing the quadratic form 
${\widetilde X}TX$ where 
\[ 
T=
\ba{cc}
\langle\hat{S}_{n1}\cdot\hat{S}_{n_{1}}\rangle &\frac{1}{2} \langle\hat{S}_{n_{2}}\cdot\hat{S}_{n_{1}}+\hat{S}_{n_{1}}\cdot\hat{S}_{n_{2}}\rangle \\
\frac{1}{2} \langle\hat{S}_{n_{2}}\cdot\hat{S}_{n_{1}}+\hat{S}_{n_{1}}\cdot\hat{S}_{n_{2}}\rangle & \langle\hat{S}_{n_{2}}
\cdot\hat{S}_{n_{2}}\rangle \\
\ea, \ \mbox{and}\ \ X=
\ba{c}
\cos\Phi \\ 
\sin\Phi
\ea
\]
It is not difficult to see that the minimum value of $(\Delta {\hat{S}}_\bot)^2$ (equivalently $\left\langle \left(\hat{S}\cdot \hat{n}_\perp\right)^2 \right\rangle$) is given by the minimum eigenvalue of the matrix $T$. Thus, one has
\begin{equation}
(\Delta {\hat{S}}_\bot)^2_{\mbox{min}}=\frac{\langle 
        \hat{S}_{n_{1}}^{2}+\hat{S}_{n_{2}}^{2}\rangle}{2}-\frac{1}{2}\sqrt{\langle{
        \hat{S}_{n_{1}}^{2}-\hat{S}_{n_{2}}^{2}\rangle}^{2}+\langle\hat{S}_{n_{2}}\cdot\hat{S}_{n_{1}}+\hat{S}_{n_{1}}\cdot\hat{S}_{n_{2}}\rangle^{2}}  
\end{equation} 
where $\hat{S}_{n_{1}}=\hat{S}\cdot \hat{n}_1$ and $\hat{S}_{n_{2}}=\hat{S}\cdot \hat{n}_2$  with $\hat{n}_1$, $\hat{n}_2$ given in 
Eq. (\ref{n1n2}).
 
A straight forward calculation gives $\langle\hat{S}_{n_{2}}\cdot\hat{S}_{n_{1}}+\hat{S}_{n_{1}}\cdot\hat{S}_{n_{2}}\rangle=0$ leading to 
	\be
	\label{minvar}
(\Delta {\hat{S}}_\bot)^2_{\mbox{min}}= \frac{\langle 
        \hat{S}_{n_{1}}^{2}+\hat{S}_{n_{2}}^{2}\rangle}{2}-\frac{1}{2}\sqrt{\langle{
        \hat{S}_{n_{1}}^{2}-\hat{S}_{n_{2}}^{2}\rangle}^{2}}  =\langle \hat{S}_{n_{2}}^{2}\rangle.
\ee
 Eq. (\ref{minvar}) provides an expression for the minimum variance of the spin component perpendicular to the mean spin direction $\hat{n}_0$. The relation $\left(\Delta(\hat{S}\cdot\hat{n}_{\perp})\right)_{\text{min}}=\left(\Delta \hat{S}_\perp\right)_{\text{min}}$ in  
Eq. (\ref{minvar}) also identifies $\hat{n}_2$ as $\hat{n}_\bot$, the direction along which the component of the spin has minimum variance. 
 
Table \ref{table03} lists the 
expressions for 
$\langle{\hat{S}_{n_{2}}^{2}\rangle}$ for all degeneracy configurations possible for $N=2$ to $N=5$.
\begin{table}[ht] 
\caption{$\langle{\hat{S}_{n_{2}}^{2}}\rangle$ for different degeneracy configurations} 
\label{table03}
\begin{tabular}{|c|l|}
\hline 
$N=2,\,k=1$ & 
$\frac{1}{2}+\frac{1}{2(1+a^{2})}\left[M_{1}M_{3}+M_{2}^{2}\right] $ 
\\ \hline 
$N=3,\, k=2$ & 
$\frac{3}{4}+\frac{1}{(1+2a^{2})}\left[\frac{1}{2}M_{1}^{2}+2M_{1}M_{2}
 a+M_{1}M_{3}+M_{2}^{2}\right]$\\ \hline
$N=3,\, k=1$ & 
$\frac{3}{4}+\frac{1}{(1+2a^{2})}\left[\frac{1}{2}M_{3}^{2}+2M_{3}M_{2}
a+M_{1}M_{3}+M_{2}^{2}\right]$\\ \hline
$N=4,\,k=2$ & 
$1+\frac{[M_{1}^{2}+4M_{1}M_{2} 
a+M_{2}^{2}a^{2}+2M_{1}M_{3}(1+a^{2})+2M_{2}^{2}(1+a^{2})+4M_{2}M_{3}a+\frac{1}{2}M_{3}^{2}]}{2(1+4a^{2}+a^{4})}$
 \\ \hline
 $N=4,\,k=1$ &
 $1+\frac{3}{(1+3a^{2})}\left[\frac{1}{2}M_{1}M_{3}+\frac{1}{2}
 M^{2}_{2}
 +2M_{2}M_{3}a+\frac{1}{2}M_{3}^{2}(1+a^{2})\right]$ \\
 \hline
  $N=5,\,k=4$ &
 $\frac{5}{4}+\frac{4}{(1+4a^{2})}\left[\frac{1}{2}M_{1}M_{3}+\frac{1}{2}
 M^{2}_{2}
 +3M_{2}M_{1}a+\frac{3}{4}M_{1}^{2}(1+2a^{2})\right]$\\  \hline
 $N=5,\,k=3$ & 
 $\frac{5}{4} + 
 \frac{\frac{3}{2}M_{1}^{2}(1+2a^{2})
 +6M_{1}M_{2}(2+a^{2})
 a+3M_{2}^{2}a^{2} +3M_{1}M_{3}(1+2a^{2})  
 +3M_{2}^{2}(1+2a^{2})+6M_{2}M_{3}a +\frac{1}{2}M_{3}^{2}}{(1+6a^{2}+3a^{4})}$ \\
  \hline
 $N=5,\,k=2$ &
 $\frac{5}{4}+
 \frac{[\frac{1}{2} M_{1}^{2}
 +6M_{1}M_{2}
 a+3M_{2}^{2}a^{2}+3M_{1}M_{3}(1+2a^{2})  
 +3M_{2}^{2}(1+2a^{2})+6M_{2}M_{3}a(2+a^{2}) +\frac{3}{2}M_{3}^{2}(1+2a^{2})]}{(1+6a^{2}+3a^{4})}$\\ 
 \hline
 $N=5,\,k=1$ &
 $\frac{5}{4}+\frac{4}{(1+4a^{2})}\left[\frac{1}{2}M_{1}M_{3}+\frac{1}{2}
 M^{2}_{2}
 +3M_{2}M_{3}a+\frac{3}{4}M_{3}^{2}(1+2a^{2})\right]$\\  \hline
\end{tabular}
\end{table} 
 In Table III,  
\begin{eqnarray} 
M_1&=&\frac{\langle \hat{S}_x \rangle}{\sqrt{\langle \hat{S}_x \rangle^2+\langle \hat{S}_z \rangle^2}} \nonumber \\
M_2&=&\frac{1}{\sqrt{\langle \hat{S}_x \rangle^2+\langle \hat{S}_z \rangle^2}}\left[ a\langle \hat{S}_x \rangle-\sqrt{1-a^{2}}\langle \hat{S}_z \rangle  \right] \\
M_3&=&\frac{1}{\sqrt{\langle \hat{S}_x \rangle^2+\langle \hat{S}_z \rangle^2}}\left[ (2a^2-1) \langle \hat{S}_x \rangle-2a\langle \hat{S}_z \rangle  \right]. 
\end{eqnarray}
 By 
careful 
analysis one can obtain a general expression for 
$\langle{ \hat{S}_{n_{2}}^{2}}\rangle$ for $N$ qubits as
\begin{eqnarray}\nonumber
\langle{ \hat{S}_{n_{2}}^{2}}\rangle & = & 
\frac{N}{4}+\frac{N(N-1)}{{\cal{N}}^{2}}  \left\{ \frac{M_{1}}{4}
\ba{c}{N-2}\\ {N-k}\ea \sum_{r=0}^{N-k}\ba{c}{k-2}\\{r} \ea  a^{2r} \left[ M_1 \ba{c}{N-k}\\ {r} \ea 
+ 2M_{2}a \ba{c}{N-k}\\ {r+1} \ea \right] \right. \nonumber\\
& & \left.+ \frac{M^{2}_{2}a^{2}}{4} \ba{c}{N-2}\\ {N-k}\ea  \sum_{r=0}^{N-k} 
\ba{c}{k-2}\\ {r} \ea \ba{c}{N-k}\\ {r+2}\ea a^{2r}+ \frac{M_{1}}{2}
\ba{c}{N-2}\\ {N-k-1}\ea \right.  \nonumber \\
& & \left. \sum_{r=0}^{N-k-1}\ba{c}{N-k-1}\\ {r} \ea a^{2r}\left[ M_2 a \ba{c}{k-1}\\ {r+1} \ea
+M_{3} \ba{c}{k-1}\\ {r} \ea \right]+
\frac{M_{2}}{2}\ba{c}{N-2}\\ {N-k-1}\ea  \right. \nonumber \\
& & \left. \sum_{r=0}^{N-k-1}\,a^{2r}
\ba{c}{k-1}\\{r}\ea \left[ M_2 \ba{c}{N-k-1}\\ {r} \ea +M_{3}a \ba{c} {N-k-1}\\ {r+1} \ea \right]+
\frac{1}{4}
\ba{c}{N-2}\\{N-k-2}\ea  \right. \nonumber \\
& & \left. \sum_{r=0}^{N-k-2}  \ba{c}{N-k-2}\\ {r}\ea a^{2r}\left[ M^2_{2}a^{2}\ba{c}{k}\\ {r+2}\ea 
+2M_{2}M_{3}a  \ba{c}{k}\\ {r+1}\ea
 +M^{2}_{3}\ba{c}{k}\\ {r}\ea \right]
\right\}
\end{eqnarray}
The spin squeezing parameter (See Eq. (\ref{xi})) now turns out to be
\begin{equation}
\xi=2\sqrt{\frac{\langle{\hat{S}^{2}_{n_{2}}}\rangle}{N}}
\end{equation} 
and can be readily evaluated for all values of $N$ and $k<N$. The variation of $\xi$ with respect to the real parameter $a$ ($0\leq a<1$) is shown in Figs 1 to 3, for different values of $N$ and $k$. 
 
 \begin{figure}[h]
	\label{n8}
	\begin{center}
		\includegraphics*[width=3in,keepaspectratio]{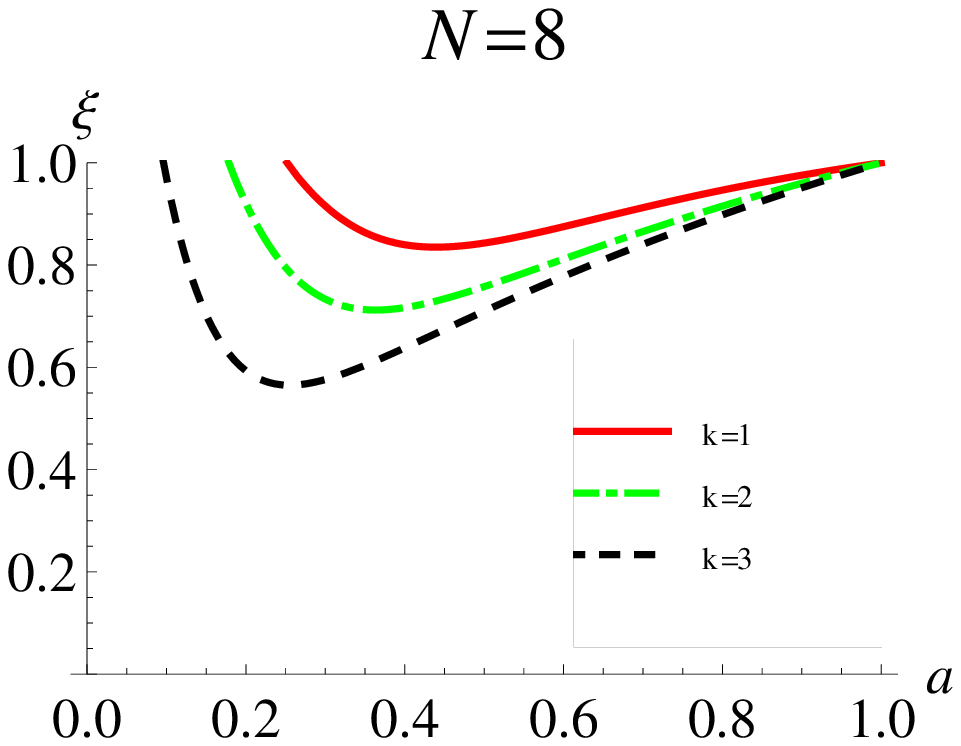}
		\includegraphics*[width=3in,keepaspectratio]{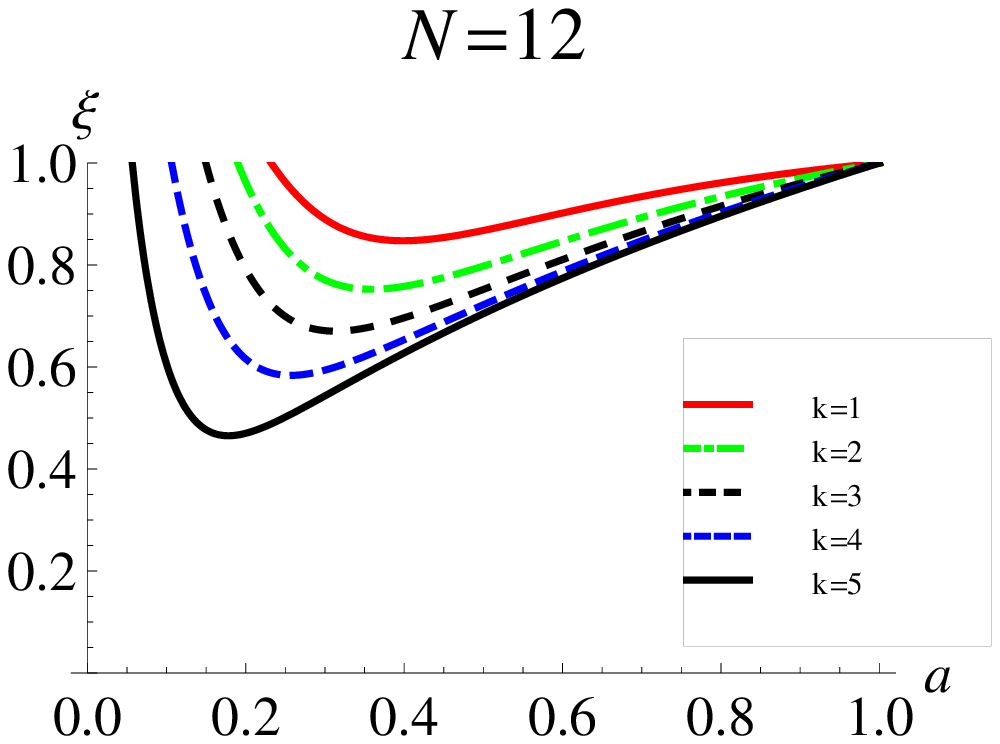}
		\caption{Variation of spin-squeezing parameter $\xi$ of the state $\vert \Psi_{k,\,N-k} \rangle$ when $N=8$ and $N=12$, for different values of $k$} 
	\end{center}
\end{figure}

\begin{figure}[h]
	\label{n100}
	\begin{center}
		\includegraphics*[width=3in,keepaspectratio]{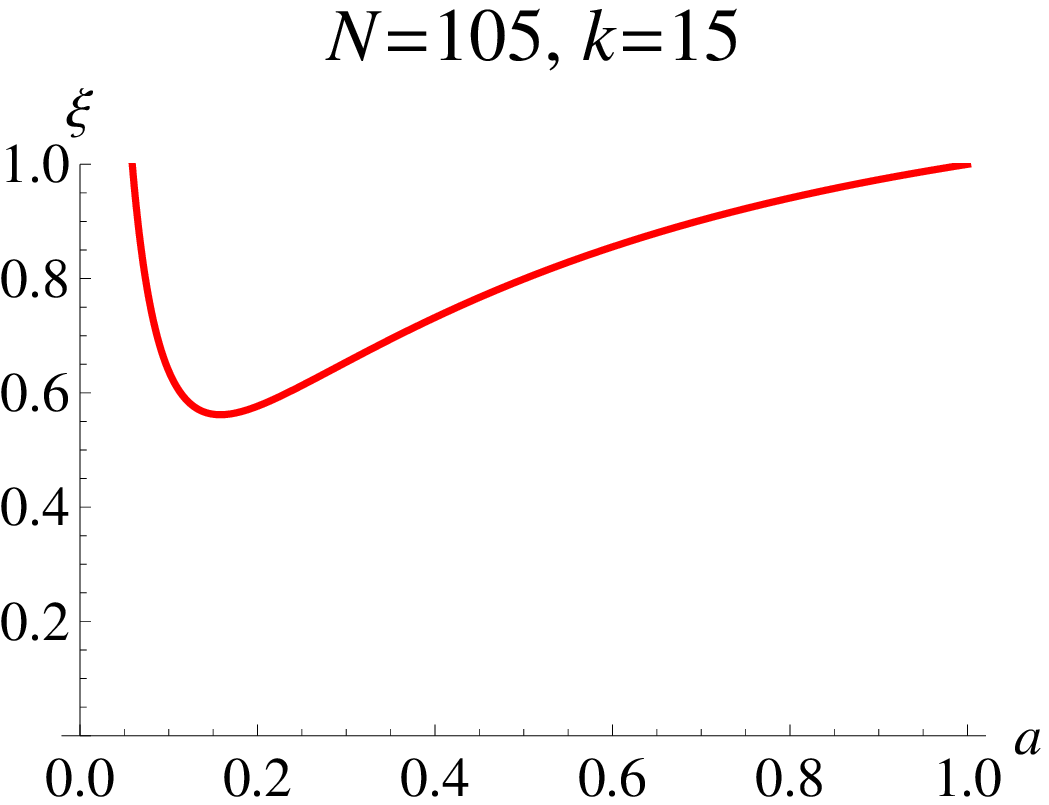}
		\includegraphics*[width=3in,keepaspectratio]{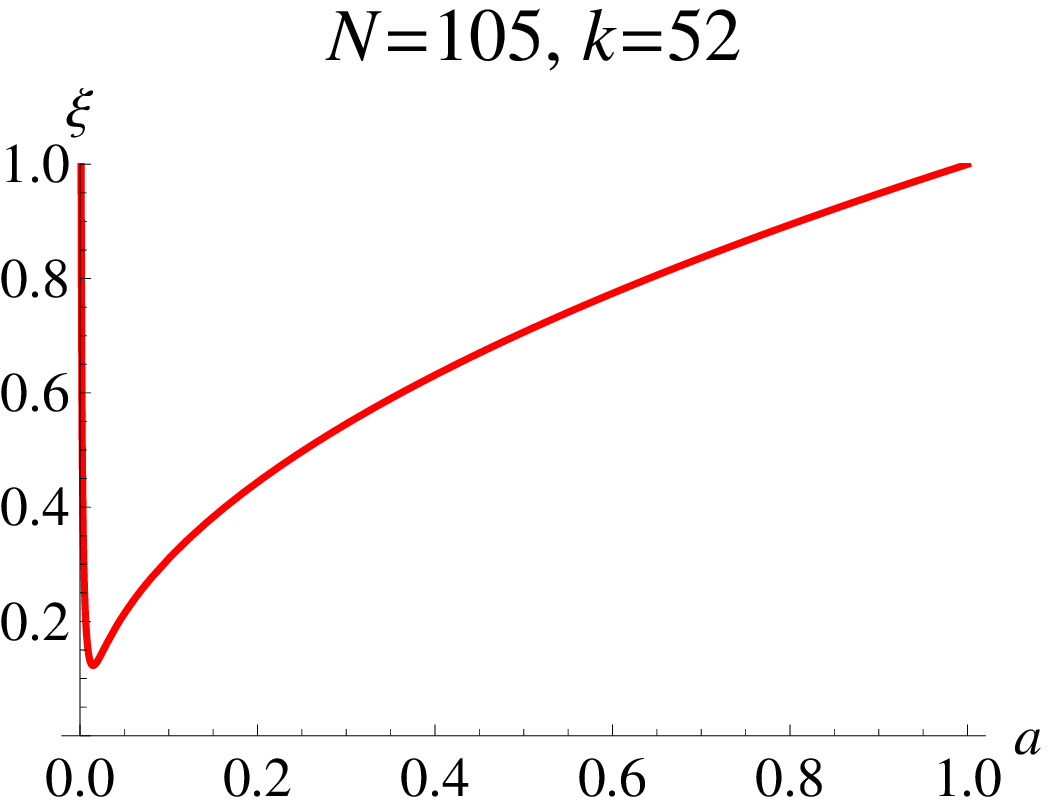}
		\caption{Comparitive plots indicating the variation of $\xi$ with respect to $k$ when $N=105$. Squeezing is seen to be more pronounced when $k=52$}  
	\end{center}
\end{figure}  

\begin{figure}[h]
	\label{n12}
	\begin{center}
		\includegraphics*[width=3in,keepaspectratio]{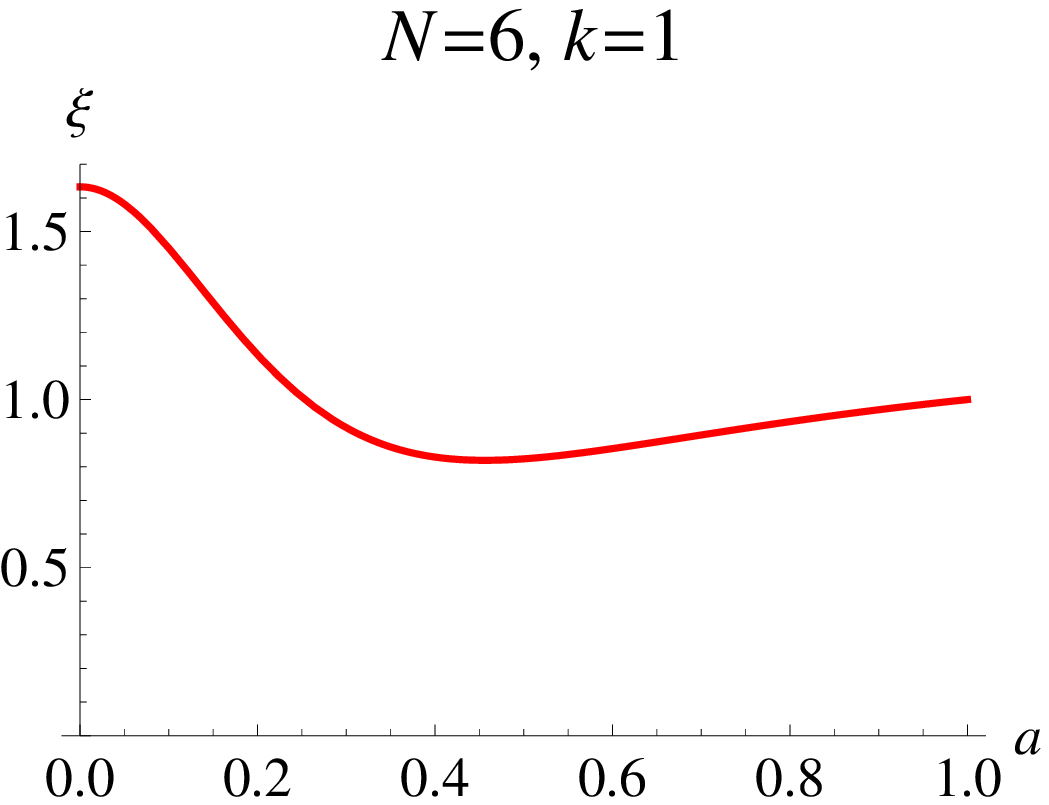}
		\includegraphics*[width=3in,keepaspectratio]{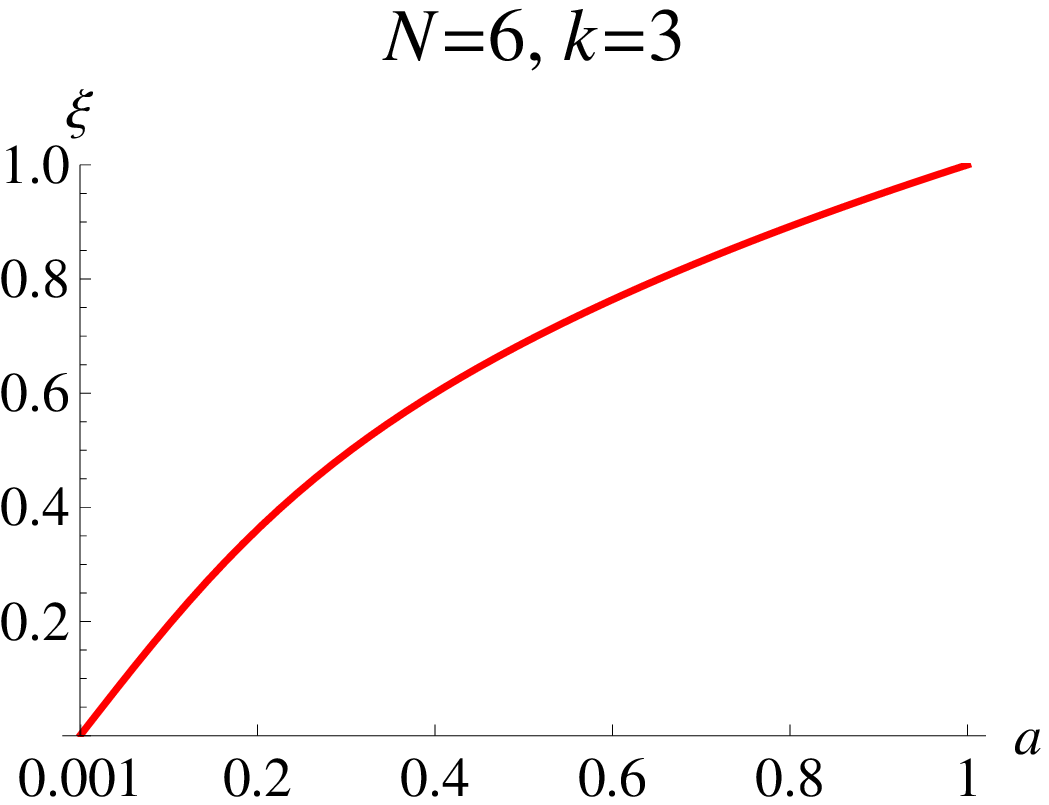}
		\caption{Variation of spin-squeezing parameter $\xi$ of the state $\vert \Psi_{1,\,4} \rangle$ and $\vert \Psi_{3,\,3} \rangle$ corresponding to
		$N=5,\ k=1$, $N=6,\ k=3$ respectively.} 
	\end{center}
\end{figure} 
 
One can draw following conclusions about the spin squeezing nature of Dicke-class of states $\vert \Psi_{k,\,N-k} \rangle$. Recall here that spin squeezing is maximum when $\xi=0$ and there is no spin squeezing when $\xi\geq 1$. 
\begin{enumerate}

\item  For any $N$, the squeezing parameter 
for the states $\vert \Psi_{k,\,N-k}\rangle$ and $\vert \Psi_{N-k,\,k}\rangle$ match with each other. This means, interchanging 
$\vert u_1\rangle$ with $\vert u_2\rangle$ in $\vert \Psi_{k,\,N-k}\rangle$ (See Eq. (\ref{maj1})) does not affect the spin-squeezing property, both of them having the same spin squeezing parameter.

\item From Figs. 1 and 2, it is evident that for a fixed $N$, squeezing increases ($\xi$ decreases) in the whole range  $0\leq a <1$ with increase in $k$ ($k<N$). But as, 
$\vert \Psi_{k,\,N-k}\rangle$ and $\vert \Psi_{N-k,\,k}\rangle$ have same $\xi$, we need to consider values of $k$ up to 
$\left[ \frac{N}{2} \right]$ where $\left[ \frac{N}{2} \right]=\frac{N}{2}$ for even $N$ and $\left[ \frac{N}{2} \right]=\frac{N-1}{2}$ for odd $N$. For instance, in Fig. 2, the graph for $N=105$, $k=15$ is equivalent to the graph for $N=105$, $k=90$. 
Also, when $\vert u_1\rangle$, $\vert u_2\rangle$ are equally distributed, i.e., when $k=\left[ \frac{N}{2} \right]$, the squeezing is pronounced in the whole range of parameter $a$, as can be seen in the graph for 
$N=105$, $k=52$.

\item The Dicke states $\vert{\Psi_{k,\,N-k}}\rangle\equiv \frac{1}{\mathcal{N}}\sum \limits_{P} 
[ \underbrace{\vert{0}\rangle \otimes \vert{0}\rangle \otimes  \ldots 
            \otimes  \vert{0}\rangle}_{k} \otimes \underbrace{\vert{1}\rangle 
            \otimes \ldots  \vert{1}\rangle}_{N-k}]$ corresponding to $a=0$ are seen to be non-squeezed (See Fig. 3). In fact, 
Dicke states with even $N$ and $k=N/2$ corresponding to equal distribution of $\vert{0}\rangle$,  $\vert{1}\rangle$, the spin squeezing parameter is undefined (See Fig. 3) as the mean spin direction becomes a null vector. Such Dicke states are the ones with $s=\frac{N}{2}$ and $m=0$. For all other Dicke states, with odd or even $N$, $k<N$, the spin squeezing parameter $\xi$ is greater than $1$ as is illustrated explicitly in the graph for $N=6,\,k=1$ (Fig. 3) and indicated in Figs. 1 and 2.  
\end{enumerate} 
One can readily conclude that Dicke-class of states $\vert \Psi_{k,\,N-k}\rangle$, containing two distinct non-orthogonal spinors (with $a\neq 0$) exhibit squeezing, the amount of squeezing varying with the degeneracy configuration $\left\{ k,\, N-k \right\}$. 	

\section{Conclusion}

Using the Majorana representation of pure symmetric 
multiqubit states with two distinct spinors, we have studied their 
spin squeezing properties. By explicit determination 
of the mean spin vector and minimum variance of the states under consideration, 
the Kitegawa-Ueda spin squeezing parameter~\cite{ku} is determined. 
Through illustrative graphs of the variation of the spin squeezing parameter, it is shown 
that symmetric multiqubit states characterized by two distinct non-orthogonal spinors 
exhibit spin-squeezing. For a fixed number of qubits in the state, the squeezing is seen 
to be maximum if the two distinct spinors characterizing the state are equal in number. 
Due to the usefulness of spin squeezed states in high precision, low noise spectroscopy 
and in quantum information theory, the study presented here is of importance. We wish to mention here that 
spin squeezing parameter of the Dicke-class of states 
considered here are determined in an alternative manner in Ref. ~\cite{aks} using the two-qubit density matrices of the symmetric multiqubit state, the results matching with that presented here.  The study of spin squeezing 
for the pure symmetric multiqubit states having diversity 
degree $3$ is under progress.

\section*{Acknowledgements}
KSA thanks the University Grant Commission for providing BSR-RFSMS 
fellowship during this work. We thank Dr. A. R. Usha Devi for 
insightful discussions.


\begin{thebibliography}{0} 
\bibitem{kim1} Kimble H J and Walls D F 1987 {\it J. Opt. Soc. B} {\bf 
4} 1450  
\bibitem{kim2} Loudon R and Knight P L 1987 {\it J. Mod. Opt.} {\bf 34}, 709 
\bibitem{kim3} Wodkiewicz K 1985 {\it Phys. Rev. B} {\bf 32}, 4750
\bibitem{ku} Kitegawa M and Ueda M 1993 {\it Phys. Rev. A} {\bf 47} 
5138   
\bibitem{wn2} Wineland D J,  Bollinger J J,  Itano W M and  Heinzen D 
J 1994
{\it Phys. Rev. A} {\bf 50} 67
\bibitem{gsapuri}  Agarwal G S and Puri R R 1990 {\it Phys. Rev. A} 
{\bf 41} 3782; 1994 \textit{ibid} {\bf 49}  4968 
\bibitem{puri}  Puri R R  1997 {\it Pramana} {\bf 48} 787
\bibitem{msd} Mallesh K S, Swarnamala Sirsi, Sbaih M A A, 
    Deepak P N and Ramachandran G 2000  {\it J. Phys. A: Math. Gen.}
    {\bf 33} 779
\bibitem{sor} Sorenson A,  Duan L M, Cirac J I and Zoller P  2001 
{\it {Nature, London}} {\bf 409} 63  
\bibitem{ulamku}  Ulam-Orgikh D and Kitagawa M 2001 {\it Phys. Rev. 
A} {\bf 64} 052106. 
\bibitem{um}  Usha Devi A R,  Mallesh K S, Mahmoud A A Sbaih,  Nalini 
K B and Ramachandran G 2003 {\it J. Phys. A: Math. Gen.} {\bf 36} 
5333 
\bibitem{usha1}  Usha Devi A R, Wang X and Sanders B C 2003 {\it 
Quantum Inf. Proc.} {\bf 2} 209.
\bibitem{usha2} A. R. Usha Devi and Sudha 2011 {\it Asian Journal of 
Physics} {\bf 20} 
\bibitem{s1}  Wang X and  Sanders B C 2003{\it Phys. Rev. A} {\bf 68} 
03382
\bibitem{s2}  Wang X and Molmer K 2002 {\it Eur.  Phys. J. D} {\bf 
18} 385 
\bibitem{us1} Usha Devi A R, Uma M S, Prabhu R and Sudha 2006 {\it 
Int. J. Mod. Phys. B.} {\bf 20} 1917 
\bibitem{us2} Usha Devi A R, Uma M S, Prabhu R and Sudha 2005 {\it J. 
Opt. B: Quantum Semiclass.Opt.} {\bf 7} S740
\bibitem{dyan} Yan D, Wang X G, Song L J and Zong Z G 2007 {\it Cent. 
Eur. J. Phys. 5} {\bf 367} 
\bibitem{akru}  Usha Devi A R,  Uma M S,  Prabhu R and Rajagopal A K 
2007 {\it Phys. Lett. A} {\bf 364} 203 
\bibitem{gtoth} Toth G, Knapp C, Guhne O, Briegel H J 2009 {\it Phys. 
Rev. A} {\bf 79} (2009) 042334. 
\bibitem{ReviewNori}  Ma J,  Wang X, Sun C P and Nori F 2011 {\it 
Phys. Rep.} {\bf 509} 89  
\bibitem{Toth2011} Vitagliano G, Hyllus P, Egusquiza I L and  
T{\'o}th G 2011 {\it Phys. Rev. Lett.} {\bf 107} 240502
\bibitem{song} Song-song Li 2011 {\it Int. J. Theor. Phys.} {\bf 50} 
719
\bibitem{divsu} Divyamani B G, Sudha, Usha Devi A R 2016  {\it Int J Theor Phys} {\bf 55}, 2324 
\bibitem{caves}  Caves C M, {\it Phys. Rev. D} {\bf 23} (1981) 1693.
\bibitem{yurke} Yurke B, {\it Phys. Rev. Lett.} {\bf 56} (1986) 1515 
\bibitem{kunew2} Kitegawa M and Ueda M, {\it Phys. Rev. Lett.} {\bf 67} (1991) 1852.
\bibitem{maj}  Majorana E 1932 {\it Nuovo Cimento} {\bf 9}  43 
\bibitem{bas} Bastin T, Krins S, Mathonet P, Godefroid M, Lamata	L and Solano 2009 {\it E Phys. Rev. Lett.} {\bf 103} 
    070503 
	\bibitem{usha4} Usha Devi A R, Sudha and  Rajgopal A K 2012 {\it 
 Quantum Inf. Proc.} {\bf 11} 685
 \bibitem{sac} Sackett C A et al 2000 {\it Nature(London)} {\bf404} 
 256 
 \bibitem{usha3} Usha Devi A R, Prabhu R and Rajagopal A K 2007 {\it 
 Phys.Rev.Lett.} {\bf98} 060501 
\bibitem{aks} Akhilesh K S, Divyamani B G, Sudha, Usha 
    Devi A R and Mallesh K S 2018 quant-ph arXiv:1803.09143v3

\end{thebibliography}
\end{document}